\documentclass[review]{elsarticle}

\usepackage{graphicx}
\usepackage{caption}
\usepackage{subcaption}
\usepackage{amsmath}
\usepackage{enumerate}
\usepackage{array}
\usepackage{float}
\usepackage{url}
\usepackage{xcolor}
\usepackage{booktabs}
\usepackage{multirow}
\usepackage{multicol} 
\newcommand{\be}{\begin{eqnarray}}
\newcommand{\ee}{\end{eqnarray}}

\usepackage{amssymb}
\usepackage{caption}
\usepackage{subcaption}

\journal{Computer Speech and Language Special Issue}

\begin{document}

\begin{frontmatter}


\title{Dereverberation of Autoregressive  Envelopes for Far-field Speech Recognition}


\author[1,2]{Anurenjan Purushothaman}
\author[2]{Anirudh Sreeram}
\author[2]{Rohit Kumar}
\author[2]{\\Sriram Ganapathy}

\address[1]{ Dept. of ECE, College of Engineering, Thiruvananthapuram, India, 695016}
\address[2]{Learning and Extraction of Acoustic Patterns (LEAP) lab, \\ Electrical Engineering, Indian Institute of Science, Bangalore, India, 560012}

\begin{abstract}
The task of speech recognition in far-field  environments is adversely affected by the reverberant artifacts that elicit as the temporal smearing of the sub-band envelopes. In this paper, we develop a neural model for speech dereverberation using the long-term sub-band envelopes of speech. The sub-band envelopes are derived using frequency domain linear prediction (FDLP) which performs an autoregressive estimation of the Hilbert envelopes. The neural dereverberation model estimates the envelope gain which when applied to reverberant signals suppresses the late reflection components in the far-field signal. The dereverberated envelopes are used for feature extraction in speech recognition. Further, the sequence of steps involved in envelope dereverberation, feature extraction and acoustic modeling for ASR can be implemented as a single neural processing pipeline which allows the joint learning of the dereverberation network and the acoustic model. Several experiments are performed on the REVERB challenge dataset, CHiME-3 dataset and VOiCES dataset. In these experiments, the joint learning of envelope dereverberation and acoustic model yields significant performance improvements over the baseline ASR system based on log-mel spectrogram as well as other past approaches for dereverberation (average relative improvements of $10$-$24$\% over the baseline system). A detailed analysis on the choice of hyper-parameters and the cost function involved in envelope dereverberation is also provided.
\end{abstract}


\end{frontmatter}


\section{Introduction}
Automatic speech recognition (ASR) is a challenging task in far-field conditions. This is particularly due to the fact that the speech signal will be reverberant and noisy. The word error rates (WER) in ASR have seen a dramatic improvement over the past decade due to the advancements in deep learning based techniques \cite{yu2016automatic}. Still the deterioration in performance in noisy and reverberant conditions persist \cite{hain2012transcribing}. A relative increase in WER of $75\%$ is reported by \cite{dan, ganapathy20183} when the signal from headset microphone is replaced with far-field array microphone signals in the ASR systems. This deterioration is due to temporal smearing of time domain envelopes caused by reverberation \cite{yoshioka2012making}.

One common approach to suppress reverberation is to combine all channels by beamforming \cite{anguera2007acoustic} before feeding it to the ASR system. Recently,  unsupervised neural mask estimator for generalized eigen-value beamforming is proposed \cite{rohit}. Traditional pre-possessing also includes the weighted prediction error (WPE)~\cite{wpe} based dereverberation along with the beamforming in most state-of-art far-field ASR systems. Further, multi-condition training is usually used to alleviate the mismatch between training and testing \cite{seltzer2013investigation}. Here, either simulated reverberant data or real far-field data can be added to the training data. However, even with these techniques, the beamformed signal shows significant amount of temporal smoothing in sub-band envelopes. The temporal smearing is caused by the superposition of direct path signal and reflected signals and this leads to ASR performance degradation~\cite{ganapathy2017multivariate}.

In this paper, we analyze the effect of reverberation on sub-band Hilbert envelopes. We show that the effect of reverberation can be approximated as convolution of the long-term sub-band envelopes of clean speech with the envelope of room impulse response function. In order to compensate for the late reverberation component in the envelope, we explore a Wiener filtering approach where the Wiener filter gain is computed using a deep neural network (DNN). The gain estimation network is implemented using a convolutional-long short term memory (CLSTM) model.  The gain is multiplied with the sub-band envelopes to suppress reverberation artifacts. The sub-band envelopes are converted to spectrographic features through integration and used for deep neural network based ASR. The sub-band envelopes are derived using the autoregressive modeling framework of frequency domain linear prediction (FDLP) \cite{thomas2008recognition,ganapathy2018far}. 

The steps involved in envelope dereverberation, feature extraction and acoustic modeling for ASR can all be implemented as neural network layers. Therefore, we also propose an approach for joint learning  of the speech dereverberation model with the ASR acoustic modeling network as a single neural model. Various ASR experiments are performed on the REVERB challenge dataset \cite{reverb} as well as the CHiME-3 dataset \cite{chime3}. In these experiments, we show that the proposed approach improves over the state-of-the-art ASR systems based on log-mel features as well as other past approaches proposed for speech dereverberation and denoising based on deep learning. In addition, we also extend the approach to large vocabulary speech recognition on VOiCES dataset \cite{voices,voices_inter}.

The rest of the paper is organized as follows. The related prior work is discussed in Section~\ref{sec:prior_work}. This section also discusses the key contributions from the proposed work. Section~\ref{sec:signal_model} provides details regarding the reverberation artifacts and autoregressive envelope estimation using frequency domain linear prediction. In Section~\ref{sec:neural_network}, we discuss the envelope dereverberation model, feature extraction as well as the joint approach to dereverberation with acoustic modeling for ASR. The ASR experiments and results are discussed in Section~\ref{sec:expt}. Various model parameter choices and additional analyses are reported in Section~\ref{sec:discuss}. This is followed by a summary of the work in Section~\ref{sec:summary}. 

\section{Related prior work}\label{sec:prior_work}
Xu et. al. in \cite{xu2014regression} attempted to find a mapping function from noisy and clean signals using supervised neural network, which is used for enhancement in the testing stage. In a similar manner, speech separation problem is also explored  with ideal ratio mask based neural mapping \cite{wang2018supervised}. Zhao et. al. proposed a LSTM model for late reflection prediction in the spectrogram domain  for reverberant speech \cite{zhao2018late}. A spectral mapping approach using the log-magnitude inputs was attempted by Han et. al \cite{han2014learning}. A mask based approach to dereverberation on the complex short-term Fourier transform domain was explored by Williamson et. al \cite{williamson2017time}.

Speech enhancement for speech recognition based on neural networks has been explored in \cite{wollmer2013feature,chen2015speech,weninger2015speech}. In Maas et. al \cite{maas2013recurrent}, a recurrent neural network is used to  map  noise-corrupted  input  features  to  their  corresponding  clean  versions. A context aware recurrent neural network based convolutional encoder-decoder architecture was used in~\cite{santos2018speech} to map the power spectral features of noisy and clean speech. In a recent work by Pandey et. al \cite{pandey_2019}, the speech enhancement is learned in the time domain itself, but using a matrix multiplication to convert the time domain signal into frequency domain and the frequency domain loss is used for training. 
This approach uses mean absolute error between the STFT frames of the clean and noisy speech for training.

The joint learning of the speech enhancement neural model and the acoustic model was attempted in \cite{wang2016joint}. Here, a DNN based speech separation model is coupled with a DNN based acoustic model and the weights are adjusted jointly.  Bo Wu et. al. \cite{wu2017unified} proposed to unify the speech enhancement neural model and the acoustic model trained separately, and then the joint model is further trained to improve the ASR performance. The power spectrum in the log domain was used as features in the enhancement stage. Bo Wu et. al.~ \cite{bowu2e2017} also explored an end-to-end deep learning approach in, where the knowledge about   reverberation time is incorporated in  DNN based dereverberation front end. This reverberation time aware-DNN enhancement module and ASR acoustic module are further trained jointly to improve the ASR cost.

The key contributions from the current work can be summarized as follows,
\begin{itemize}
    \item Deriving a signal model for reverberation effects on sub-band speech envelopes and posing the dereverberation problem as a gain estimation problem.
    \item Dereverberation of the autoregressive estimates of the sub-band envelope using a CLSTM model followed by feature extraction for ASR.
    \item Joint learning of the dereverberation model parameters and the acoustic model for ASR in a single neural pipeline. 
    \item Illustrating the performance benefits of the proposed approach for multiple ASR tasks.
\end{itemize}
We use FDLP features  \cite{anu1} for far-field speech. This paper extends the prior work done in \cite{inter2020cleaning}  by proposing a joint neural dereverberation which forms an elegant neural learning framework. Further, several ASR experiments with the joint modeling approach are also conducted in this work. 

\section{Sub-band Envelopes - Effect of Reverberation and Autoregressive Estimation}\label{sec:signal_model}
We present the signal model for reverberation and the autoregressive model for estimating the sub-band envelopes~\cite{ganapathy2017multivariate,ganapathy2009autoregressive}. 
\subsection{Signal model}
When speech is recorded in far-field reverberant environment, the data collected in the microphone is modeled as 
\be
\label{eq:reverb_sig}
r(t) = x(t)*h(t),
\ee
where $x(t)$, $h(t)$ and $r(t)$  denote the clean speech signal, the room impulse response and the reverberant speech respectively. The room response function $h(t) = h_e(t) + h_l(t)$, where $h_e(t)$ and $h_l(t)$ represent the early and late reflection components. 

\textcolor{black}{Let $\textcolor{black}{x_q(n)}$, $\textcolor{black}{h_q(n)}$ and $\textcolor{black}{r_q(n)}$ denote the decimated sub-band clean speech, room-response and the reverberant speech signal respectively. Here $q=1,..,Q$ denotes the sub-band index and $n$ denotes the decimated time-index (frame)}. Assuming an ideal band-pass filtering we can write (using Eq.~\ref{eq:reverb_sig}),
\be
\textcolor{black}{r_q(n) = x_q(n)*h_q(n)}
\ee
\textcolor{black}{In the proposed model, we explore the modeling of the sub-band temporal envelopes. In order to extract the envelopes, the analytic signal based demodulation is proposed. The analytic representation of a real-valued signal is the complex   signal consisting of the original function (real part) and the Hilbert transform (imaginary part). The negative frequency components of the analytic signal are zero-valued. By representing the real-valued functions in analytic domain, the extraction of the modulation components (like envelopes and carrier signals) is facilitated. }

\textcolor{black}{
Now, the analytic signal of the sub-band signal $\textcolor{black}{r_q(n)}$ is denoted as 
$\textcolor{black}{r_{aq}(n)}$,  where 
$\textcolor{black}{r_{aq}(n) = r_q(n) + j \mathcal {H} [r_q(n)}]$. Here, $\mathcal{H} [.]$ is the Hilbert operator. It can be shown that} \cite{thomas2008recognition,ganapathy2012temporal},
\be
\label{eq:reverb_analytic_conv}
\textcolor{black}{r_{aq}(n) = \frac{1}{2}[x_{aq}(n)*h_{aq}(n)]},
\ee
\textcolor{black}{ If two signals have a modulating envelope on the same modulating sinusoidal carrier signal (single AM-FM signal), the convolution operation of the two signals will have an envelope which is the convolution of the two envelopes, i.e., the envelope of the convolution of the two signals is the convolution of the envelope of the signals. For sub-band speech signals, this envelope convolution model will form a good approximation if the sub-band signals are narrow-band. }   

Then, for band-pass filters with narrow band-width, we get the following approximation between the sub-band envelope (defined as the magnitude of the analytic signal) components of the reverberant signal and those of the clean speech signal.
\be
\label{eq:envelope_conv_model}
\textcolor{black}{m_{rq}(n) \simeq \frac{1}{2} m_{xq}(n)*m_{hq}(n)},
\ee

where $m_{rq}(n)$, $m_{xq}(n)$, $m_{hq}(n)$ denote the sub-band envelopes of reverberant speech, clean speech and room response respectively. We can further split the envelope into  early and late reflection coefficients. 
\be
\label{eq:envelope_conv_model_early}
\textcolor{black}{m_{rq}(n) = m_{rqe}(n) + m_{rql}(n)},
\ee
\subsection{Autoregressive modeling of sub-band envelopes}
Frequency domain linear prediction (FDLP) is the frequency domain dual of the conventional time domain Linear Prediction (TDLP). Just as the TDLP estimates the spectral envelope of a signal, FDLP estimates the temporal envelope of the signal~\cite{ganapathy}, i.e. square of its Hilbert envelope \cite{analytic}. The Hilbert envelope is given by the inverse Fourier transform of the auto-correlation function of discrete cosine transform (DCT)~\cite{ganapathy2012signal,athineos2003frequency}. 

We use the auto-correlation of the DCT coefficients to model the temporal envelope of the signal. The autoregressive (AR) modeling property of linear prediction implies that the model preserves the peak location of the signal (which tend to be more robust in the presence of noise and reverberation)~\cite{ganapathy2014robust}. For the FDLP model,  the sub-band AR model tries to preserve the peaks in temporal envelope~\cite{ganapathy2009autoregressive}.

Let $x(t)$ denote an $N$-point discrete sequence. The type-I odd DCT \cite{tdct} $y[k]$ for $k = 0,1,\ldots, N-1$ is given by,
\begin{equation}
y[k] =\sum_{t=0}^{N-1}c_{t,k}x(t)cos\left(\frac{2\pi tk}{M}\right)
\end{equation}
where $c_{t,k}=1$ for $t,k>0$ and $c_{t,k}=\frac{1}{2}$ for $t,k=0$ and  $c_{t,k}=\frac{1}{\sqrt{2}}$  for the values of $t, k$ where only one of the index is $0$ and $M=2N-1$. 

\textcolor{black}{
An even symmetric version of the input signal $x(t)$ is the signal $q(t)$ of length $M=2N-1$ where,}
\begin{eqnarray} 
\color{black}{
q(t) = x(t), ~~t=0..N-1 ~~~}  \nonumber \\ 
\color{black}{
q(t) = x(M-t), ~~t=N..M} \nonumber \end{eqnarray} 
\textcolor{black} {The analytic signal of a discrete time sequence can be defined using the one-sided discrete Fourier transform (DFT)~\cite{ganapathy2012signal}. The analytic signal $q_a(t)$ of the even-symmetric signal $q(t)$ can be shown to be~\cite{ganapathy2009autoregressive} the zero-padded DCT (upto scale)   $\hat{y}[k]$, where  $\hat{y}[k] = y[k],k=0..N-1$ and $\hat{y}[k] =0,k=N,..,M$.  
}

Further, it can be shown that~\cite{ganapathy2012signal},
the auto-correlation of the zero-padded DCT signal $\hat{y}[k]$ and the squared magnitude of the analytic signal (Hilbert envelope) of the even-symmetric signal $|q_a(t)|^2$ are Fourier transform pairs~\cite{ganapathy}. Hence, the application of linear prediction on the zero-padded DCT signal yields the AR model of the Hilbert envelope of signal.

\textcolor{black}{Let the linear prediction coefficients obtained from the zero-padded DCT signal be denoted as $\{a_k\}_{k=0}^p$, where $p$ is the order of the LP. The FDLP model for the envelope is given by, }
\begin{eqnarray}
\color{black}{E(n) = \frac {\sigma}{|\sum_{k=0}^p a_k e^{-2 \pi i k n} |^2} }  
\end{eqnarray} 
\textcolor{black}{where $\sigma$ denotes the LP gain. The envelope estimated in above equation represents the autoregressive model of the temporal envelopes. Note that, when the model is applied on sub-band DCT coefficients, the envelope estimated will be the sub-band temporal envelope. 
}

In this work, the sub-band envelopes of speech in mel-spaced bands are estimated using FDLP. Specifically, the discrete cosine transform (DCT) of sub-band signal $r_q(t)$ is computed and a linear prediction (LP) is applied on the DCT components. The LP envelope estimated using the prediction on the DCT components provides an all-pole model of the sub-band envelopes $m_{rq}(n)$.
\begin{figure*}[t!]
  \centering
  \includegraphics[scale=.295]{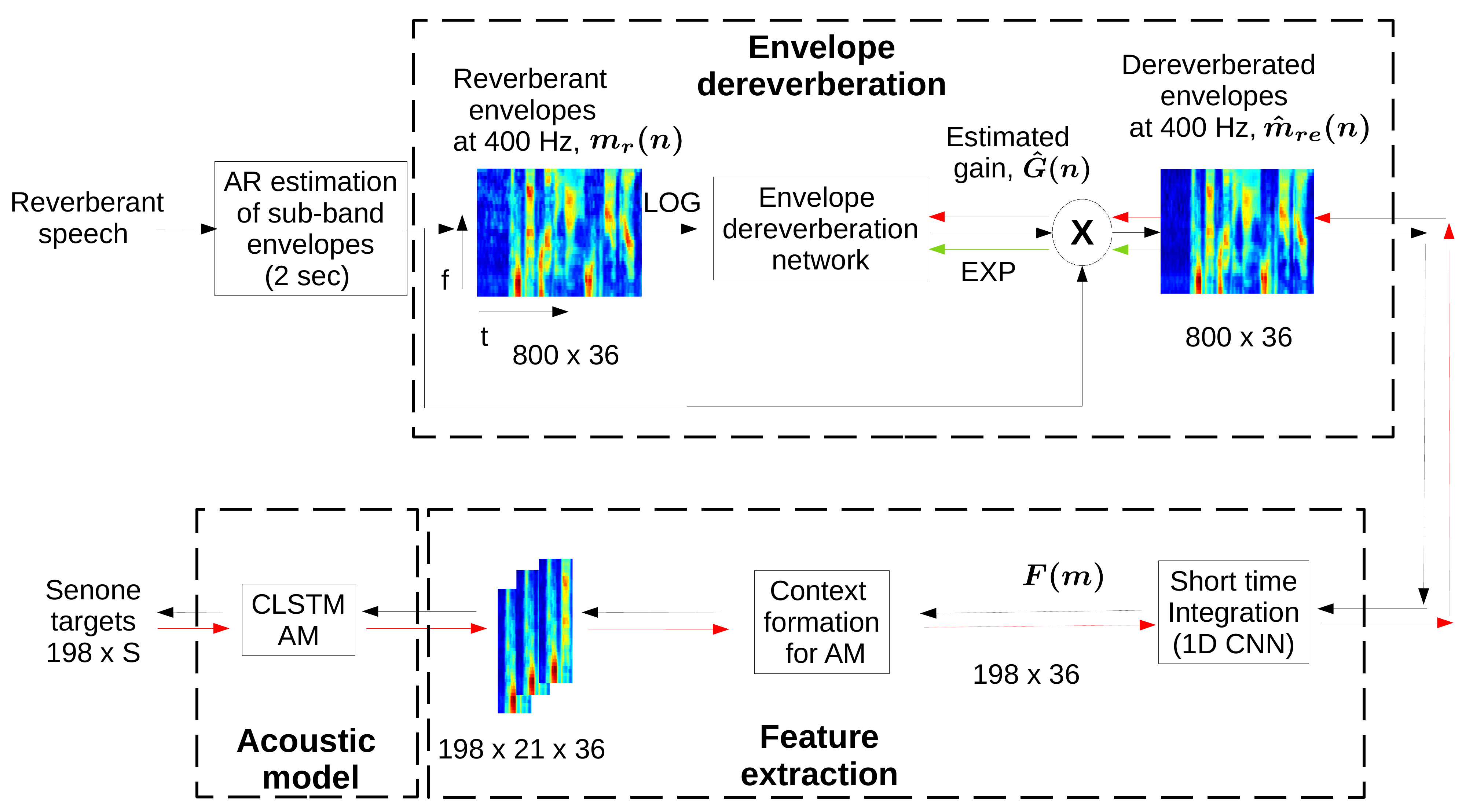}
      \caption{Block schematic of envelope dereverberation model, the feature extraction module and acoustic model (Here $m_r(n)$, $\hat{G}(n)$, $\Hat{m}_{re}(n)$ and $F(m)$ are given by Eq.s~(5, 8-11), the subscript $q$ is dropped to indicate the fact that, signals from all the bands are considered). The entire model can be constructed as an end-to-end neural framework. The black arrows denote the forward pass, the red arrows represent backward propagation with ASR loss ($E_{CE}$), and green arrows denote the backward propagation with mean square error loss ($E_{MSE}$). Here, $S$ is the total number of senone targets.}
  \label{fig1}
\end{figure*}
\section{Envelope Dereverberation and Joint Modeling}\label{sec:neural_network} 
The proposed framework (Figure \ref{fig1}), consists of three modules, (\romannumeral 1) envelope dereverberation, (\romannumeral 2) feature extraction and (\romannumeral 3) ASR acoustic model.  

\subsection{Neural dereverberation network}
As seen in Eq.~(\ref{eq:envelope_conv_model_early}), the FDLP envelope of reverberant speech can be expressed as sum of the direct component (early reflection) and those with the late reflection. In the envelope dereverberation model, our aim is to input the envelope of the reverberant sub-band temporal envelope $\textcolor{black}{m_{rq}(n)}$ to predict the late reflection components $\textcolor{black}{m_{rql}(n)}$. Once this prediction is achieved, the late reflection component can be subtracted from the sub-band envelope to suppress the artifacts of reverberation. A similar analogy to this envelope subtraction approach is the spectral subtraction model where the noise and clean power spectral density (PSD) gets added in noisy speech PSD. If Gaussian assumptions are made for PSD components \cite{martin2005speech}, the Wiener filtering approach to noisy speech enhancement provides the minimum mean squared error, where the noisy PSD is multiplied by the gain of the filter. In a similar manner,  we pose the dereverberation problem as an envelope gain estimation problem.

\textcolor{black}{The envelope gain ($G_{q}$) is defined as},

\begin{eqnarray}
\color{black}{G_q(n) = \frac{\hat{m}_{rqe}(n)}{\hat{m}_{rqe}(n) + \hat{m}_{rql}(n)} } 
\end{eqnarray}

\textcolor{black}{
The gain $G_q(n)$ is estimated using the input sub-band envelope $m_{rq}(n)$. With the gain estimate, the dereverberated envelope can be computed as, }
\begin{eqnarray}
\color{black}{\hat m_{rqe}(n) = G_q(n) m_{rq}(n)}  
\end{eqnarray}

\textcolor{black}{
The product model of enhancement is inspired by Wiener filtering principles. This sub-band envelope gain estimation is achieved using a deep neural network model in the proposed work. Following the model training, the dereverberation is achieved by multiplying the estimated sub-band envelope gain with the sub-band envelope of reverberant speech.}

The block schematic of the envelope dereverberation model is shown in Figure~\ref{fig1}. The input to the dereverberation model is the FDLP sub-band envelope of the reverberant speech. The model is trained to learn the sub-band envelope gain which is the ratio of the clean envelopes (direct component) with the reverberant envelopes. \textcolor{black}{During the model training, the model inputs are either far-field microphone recordings or the simulated reverberant recordings. The model targets are the envelope gain (Eq 8) computed using either the close talking/near-room microphone corresponding to the far-field microphone data, or  the   clean close-talking microphone data for the simulated reverberant training data. Thus, model is trained with paired data to estimate the gain.}

As the envelopes and the gain parameters are positive in nature, the model implementation in the neural architecture uses a logarithmic transform at the input and the estimated gain is transformed by an exponential operation.  \textcolor{black}{Specifically, the input to the dereverberation model is the set of sub-band envelopes $\{ log (m_{rq})(n)\} _{q=1}^Q$, where $Q$ is the number of sub-bands. The model is trained to predict the log-gain $\{ log (G _q)\} _{q=1}^Q$. The sub-band dereverberated envelope is, }
\begin{eqnarray}\label{eqn:deverberation}
\color {black}{\hat {m}_{rqe} = exp \big [ (log ({\hat {G}_q(n)) + log (m_{rq}(n)) \big ]}}
\end{eqnarray} 
\textcolor{black}{where $\hat {G}_q(n)$ is the estimate of the gain from the model. } 

The entire model developed in Section~\ref{sec:signal_model} is applicable only on long analysis windows (which are typically greater than the T60 of the room response function). Hence,  the proposed approach operates on long temporal envelopes of the order of $2$ sec. duration.    
From the reverberant speech and the corresponding clean speech, the FDLP sub-band envelopes corresponding to $2$ sec. non-overlapping segments are extracted. If the input sampling rate is $16$ kHz, a $2$ sec. segment will correspond to $32$k samples ($\textcolor{black}{t=\{1..32000\}}$). The FDLP envelopes are extracted at a sampling rate of $400$ Hz. Thus, $2$ sec. segment of audio corresponds to $800$ envelope samples ($\textcolor{black}{n=\{1..800\}}$)   for each sub-band.

The input $2$-D data of sub-band envelopes ($800$ samples from $36$ mel sub-bands) are fed to a set of convolutional layers where the first two layers have $32$ filters each with kernels of size of $41 \times 5$. The next two CNN layers have $64$ filters with $21 \times 3$ kernel size. All the CNN layer outputs with ReLU activations are zero padded to preserve the input size and no pooling operation is performed.  The output of the CNN layers are reshaped to perform time domain recurrence using $3$ layers of LSTM cells. The first two LSTM layers have $1024$ cells while the last layer has $36$ cells corresponding to the size of the target signal (envelope gain). The training criteria is based on the mean square error between the target and predicted output. The model is trained with stochastic gradient descent using Adam optimizer~\cite{adam}.

\subsection{Feature Extraction and Acoustic Modeling}
For feature extraction, the sub-band envelopes are integrated in short Hamming shaped windows of size $25$ ms with a shift of $10$ ms \cite{ganapathy2012signal}. A $25$ ms slice corresponds at $400$ Hz sampling (FDLP envelopes are sampled at $400$ Hz) to $10$ samples and the hop size of $10$ ms corresponds to $4$ samples. 

The windowed FDLP envelopes are multiplied with a Hamming shaped  window (size of $10$) and accumulated. This window is shifted by $4$ samples. A log compression is applied to limit the dynamic range of values. Following this integration, a $2$ sec. chunk of $800 \times 36$ sub-band FDLP envelopes becomes $198 \times 36$.

\textcolor{black}{In particular, let $\hat{m} _{rqe}(n)$ denote the dereverberated sub-band envelope obtained using Eq.~(\ref{eqn:deverberation}). Further, let $w(n)$ denote a Hamming window of size $10$   (corresponding to $25$ ms at $400$Hz sampling). Then, the features for ASR are extracted as, }
\begin{eqnarray}
\color{black}{
F_q(m) = log (\hat{m} _{rqe}(m) * w(m)) }
\end{eqnarray}
\textcolor{black}{where $*$ is the convolution operation, and $F_q$ denotes the scalar feature of $q$th sub-band.  Here, $m$ denotes the feature frame index at $10$ms sampling ($100$ Hz). The features for all the $Q$ sub-bands are spliced to form the final feature vector for ASR model training. }

The set of operations described above for short-term integration can  be implemented as a $1$-D CNN layer with a fixed Hamming shaped kernel size of $10$ and a stride $4$. A log non-linearity is applied on the convolution output. 

The integrated envelopes are used as time-frequency representations for ASR training. A context of $21$ frames, with $10$ frames on the left and $10$ frames on the right is used in the acoustic model training. 

\subsection{Acoustic Model}
The architecture of the acoustic model is based on convolutional long short term memory (CLSTM) networks (Figure~\ref{fig1}). The acoustic model corresponds to 2-D CLSTM network described in \cite{anu1}, consisting of $4$ layers of CNN, a layer of LSTM with $1024$ units performing recurrence over frequency and $3$ fully connected layers with batch normalization.

\subsection{Joint learning}
As shown in Figure~\ref{fig1}, the three modules of (i) envelope dereverberation, (ii) feature extraction and context formation and (iii) the ASR acoustic modeling can be combined into a single neural end-to-end framework\footnote{\textcolor{black}{The implementation of the work can be found in \url{https://github.com/iiscleap/FDLP_Envelope_Dereverberation}}}. The intermediate envelope integration step is  implemented as a $1$-layer of $1$-D convolutions with Hamming shaped kernel and log non-linearity. The context creation for acoustic features in the given  segment is also performed as a fixed $1$-D convolution layer. In this manner, the entire processing pipeline can be performed using an elegant joint learning approach. 

For generating mini-batches in the model training, a $2$ sec. speech segment is read along with the corresponding frame level targets  ($198$ frames of senone targets for the $2$ sec. segment). The entire joint neural network is trained using a combination of ASR cross entropy training criterion and mean square loss between the clean and reverberant envelopes. The gradients from the ASR loss at the input of the acoustic model (computed for each senone target) is accumulated over all the frames in the given $2$ sec. segment. This accumulated gradient is of size $198\times36$ which corresponds to the size of the integrated envelopes. This ASR loss function when further back-propagated through fixed 1-D CNN layer provides a gradient matrix of size $800\times36$. The gradient w.r.t. mean square error (MSE) between the target envelopes and the dereverberation model outputs is combined with the ASR based gradient for  training the joint model. The two gradients are indicated by two different backward arrows in Figure~\ref{fig1}.

\subsubsection{Joint loss function}
\begin{figure*}[t!]
     \centering
     \begin{subfigure}[b]{0.45\textwidth}
         \centering
         \includegraphics[width=\textwidth]{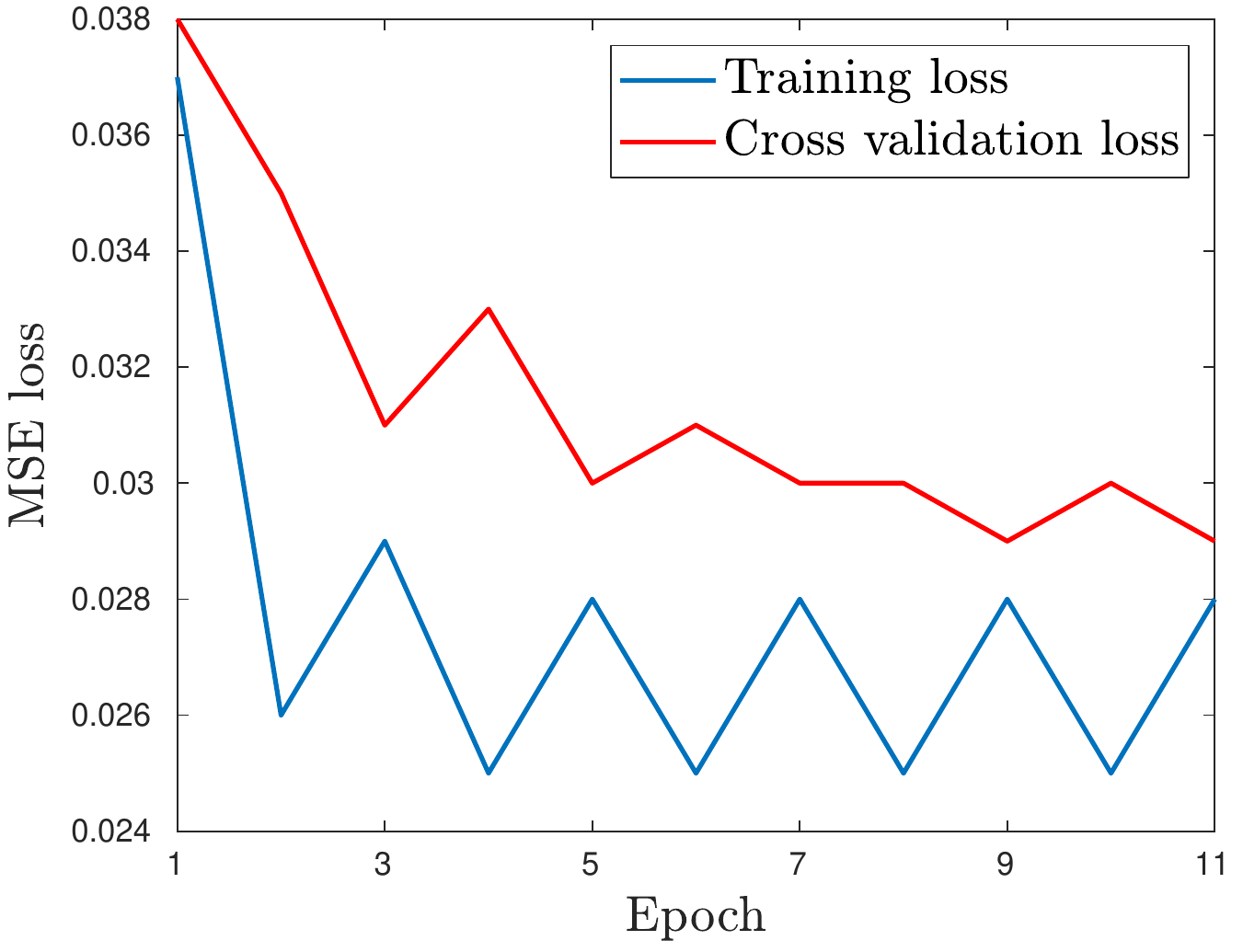}
         \caption{MSE loss vs epoch plot - dereverberation model training}
         \label{fig:y equals x}
     \end{subfigure}
     \hfill
     \begin{subfigure}[b]{0.45\textwidth}
         \centering
         \includegraphics[width=\textwidth]{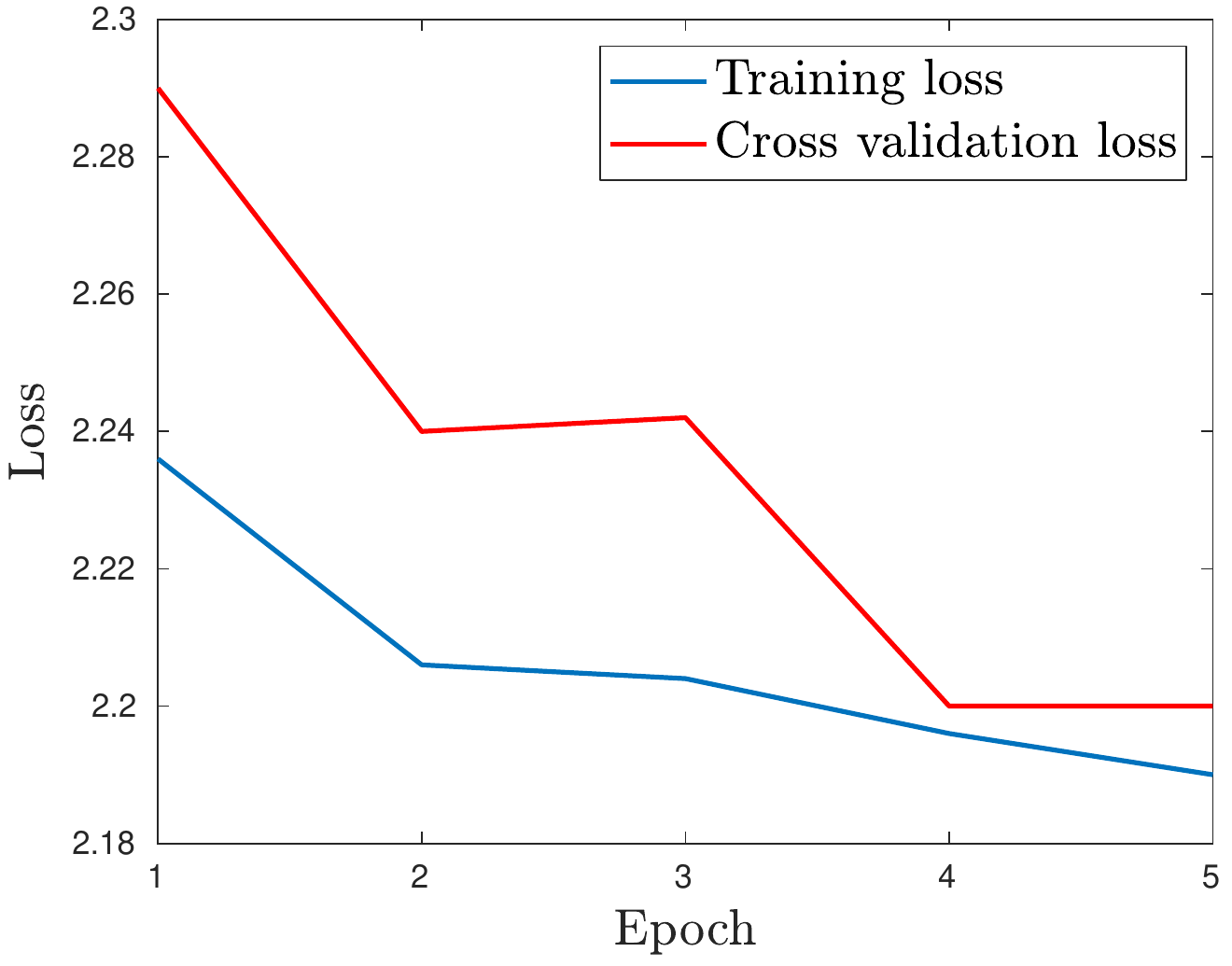}
         \caption{Loss vs epoch plot - joint model training\\}
         \label{fig:three sin x}
     \end{subfigure}
     \caption{Variation of training loss and cross validation loss with training epochs. The envelope dereverberation model and the ASR model are pre-trained  before the joint learning step.}
        \label{fig:three graphs}
\end{figure*}

The separate deverberation model is trained to minimize the mean square error loss, $E_{MSE}$, which is the squared error between the reverberant envelope and the clean counter part. For joint training, we have two loss functions, one is the mean square error loss, $E_{MSE}$  for a mini-batch and the cross-entropy loss, $E_{CE}$ between the senone targets and the corresponding posteriors for the same mini-batch. We use a combination of these two losses. Thus the final joint loss, $E_{Total}$ is given by,
\be
\label{eq:joint_loss}
E_{Total} = E_{CE}  + \mu\times E_{MSE},
\ee
where $\mu$ is a regularization parameter, which decides the share of $E_{MSE}$ in the joint loss, $E_{Total}$. In all our ASR experiments, we have used regularization parameter $\mu = 0.4$. \textcolor{black}{The absolute value of the two loss functions (different dynamic range in Figure~\ref{fig:three graphs})  does not have an impact as the model is trained with the gradient of the losses. \textcolor{black}{Note that the MSE loss changes by $0.014$ over the course of joint training while the combined loss changes by $0.04$, so they generally have comparable dynamic ranges, even if these occur at different offsets from $0$.} The regularization constant $\mu$ controls the trade-off between the two loss functions. }

The variation of the MSE loss in the envelope dereveberation network is shown in Figure~\ref{fig:three graphs}. The joint loss function on the training and validation data is also shown in this Figure. While the MSE loss trained with a higher learning rate exhibits oscillatory behavior, the  joint loss function is relatively smooth. The final joint model is used in our ASR experiments. 
\begin{figure*}[t!]
  \centering
  \includegraphics[scale=.42]{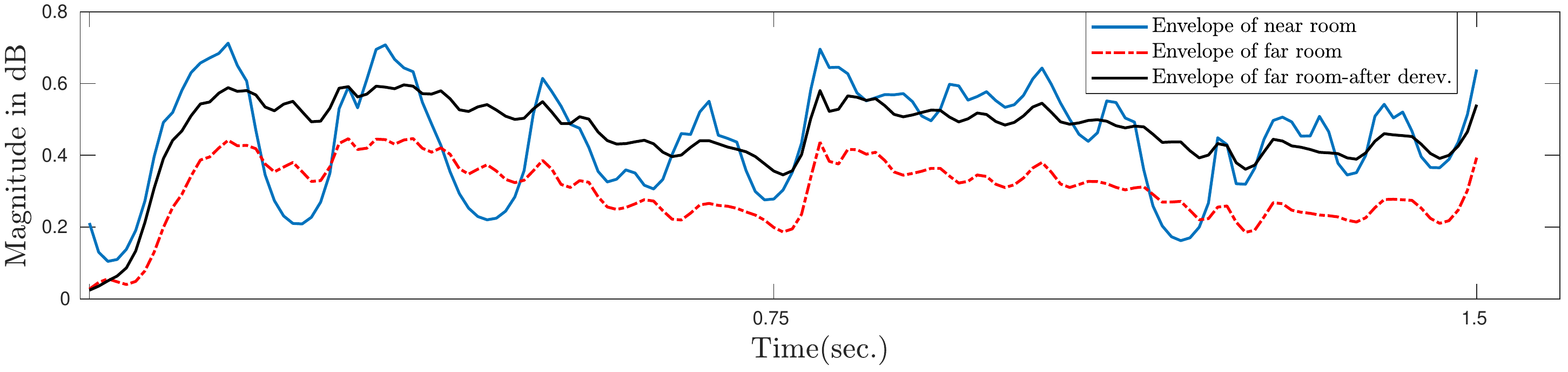}
      \caption{Comparison of temporal envelopes, FDLP envelopes for clean, reverberant speech and reverberant speech after proposed joint learning based dereverberation, recordings from the REVERB Challenge dataset.}
      \vspace{-0.3cm}
  \label{fig:env_enhancement}
\end{figure*}

\textcolor{black}{A visualization of the dereverberation, achieved for the sub-band envelope of one single sub-band ($10$ th mel-band), is shown in Figure~\ref{fig:env_enhancement}. The sub-band envelopes of reverberant signal deviate from their clean signal counterparts (as explained in Sec.~\ref{sec:signal_model}). Using the dereverberation model proposed in this paper, we find that the FDLP envelopes are more closely matched with the clean signal envelopes.}  In Section~\ref{sec:expt}, we compare the performance of the CLSTM acoustic model architecture with other model architectures for dereverberation.

\section{Experiments and results}\label{sec:expt}
The experiments are performed on REVERB challenge \cite{reverb} and CHiME-3 \cite{chime3} datasets. For the baseline model, we use WPE enhancement \cite{wpe} along with unsupervised GEV beamforming \cite{rohit}. This signal is processed with filter-bank energy features (denoted as BF-FBANK). The FBANK features are $36$ band log-mel spectrogram with frequency range of $200-6500$ Hz. This is the same frequency decomposition used in the FDLP and FDLP-dereverberation experiments. The acoustic model is the  2-D CLSTM network described in \cite{anu1}.
\begin{table}[t!]
\caption{Word Error Rate (\%) in REVERB dataset for different features and proposed dereverberation method. Here prop denotes proposed work in this paper.}
\resizebox{\columnwidth}{!}{
\begin{tabular}{@{}l|ccc|ccc@{}}
\toprule
\multicolumn{1}{c|}{\multirow{2}{*}{\textbf{\begin{tabular}[c]{@{}c@{}}Model\\ Features\end{tabular}}}} & \multicolumn{3}{c|}{\textbf{Dev}}                                              & \multicolumn{3}{c}{\textbf{Eval}}                                             \\ \cmidrule(l){2-7} 
\multicolumn{1}{c|}{}                                                                                   & \multicolumn{1}{l}{\textbf{Real}} & \multicolumn{1}{l}{\textbf{Simu}} & \multicolumn{1}{l|}{\textbf{Avg}} & \multicolumn{1}{l}{\textbf{Real}} & \multicolumn{1}{l}{\textbf{Simu}} & \multicolumn{1}{l}{\textbf{Avg}} \\ \midrule
BF-FBANK                                                                                          & 19.1                     & 6.1                      & 12.6                     & 14.7                     & \textbf{6.5}                      & 10.6                    \\
BF-FDLP ~\cite{anu1}                                                                                          & 17.8                     & 6.8                      & 12.3                     & 14.0                     & 7.0                      & 10.5                    \\
BF-FBANK + CLSTM derevb. (prop)                                                                                            & 17.3                     & 5.5                      & 11.4                     & 13.1                     & 6.9                      & 10.0                    \\
BF-FBANK + spectral mapping derevb.\cite{spec_map}                                                                                            & 15.8                     & \textbf{5.2}                      & 10.5                     & 12.8                     & 6.7                      & 9.7                    \\
BF-FBANK + context aware derevb.\cite{santos2018speech}                                                                                            & 19.6                     & 6.9                      & 13.2                     & 17.5                     & 9.0                      & 13.2                    \\
BF-FBANK + end to end derevb.\cite{bowu2e2017}                                                                                            & -                     & -                      & -                     & 24.8                     & 7.9                      & 16.4                    \\
BF-FDLP + CLSTM derevb. (prop)                                                                                           & 16.3          
          & 5.6                      & 10.9                      & 13.4                    & 7.1                      & 10.2                     \\
BF-FDLP + CLSTM derevb.  + joint (prop)                                                                                           & \textbf{15.2 }                    & 5.6                     & \textbf{10.4}                      & \textbf{12.1}                     & 7.1                      & \textbf{9.6}                     \\ \bottomrule
\end{tabular}}
\label{table:1}
\end{table}
\subsection{ASR framework}
  We use the Kaldi toolkit \cite{kaldi} for deriving the senone alignments used in the PyTorch deep learning framework for acoustic modeling. A hidden Markov model - Gaussian mixture model (HMM-GMM) system is trained with MFCC (Mel Frequency Cepstral Coefficients) features \cite{mfcc} to generate the alignments for training the CLSTM acoustic model. A tri-gram language model \cite{trigram} is used in the ASR decoding and the best language model weight  obtained from development set is used for the  evaluation set.
  \subsection{REVERB Challenge ASR}
The REVERB challenge dataset \cite{rev3} for ASR consists of $8$ channel recordings with real and simulated reverberation conditions. The  simulated data is comprised of reverberant utterances generated (from the WSJCAM0 corpus \cite{rev1}) by artificially convolving clean WSJCAM0 recordings with the measured room impulse responses (RIRs) and adding noise at an SNR of $20$ dB. The simulated data has six different reverberation conditions. The real data, which is comprised of utterances from the MC-WSJ-AV corpus \cite{rev2}, consists of utterances spoken by human speakers in a noisy reverberant room. The training set consists of $7861$ utterances from the clean WSJCAM0 training data by convolved with $24$ measured RIRs.
\subsubsection{Discussion}
Table \ref{table:1} shows the WER results for experiments on REVERB challenge dataset. The WPE along with unsupervised GEV beamformed signal is used for  all the ASR experiments (denoted as BF). The BF-FDLP baseline by itself is better than the BF-FBANK baseline (average relative improvements of $2$\% on the development set and about $1$\% on the evaluation set). For a fair comparision of the proposed approach, we have applied a similar dereverbaration method on BF-FBANK baseline. Here, we have trained the neural model with log-mel features corresponding to $2$ sec. duration with all the $36$ mel-bands jointly. This approach is denoted as BF-FBANK + CLSTM derevb. (prop). Average relative improvements of $10$\% on the development set and about $6$\% on the evaluation set is achieved compared to the BF-FBANK baseline. 

BF-FBANK + spectral mapping derevb. \cite{spec_map} corresponds to the work by Kun Han et. al. Here a $3$-layer deep neural network of $2570$ units is used as the dereverberation neural model. The network is fed with $257$-dimensional log-magnitude STFT features from a frame of $25$ m.sec. A context window of $10$-frames ($5$-left and $5$-right) is selected and the network tries to predict the central frame. The work by Santos et. al. is implemented as BF-FBANK + context aware derevb.~\cite{santos2018speech}. A CNN-GRU based encoder-decoder model sees the entire utterance at the $257$-STFT magnitude level features and trained to predict the clean utterance. The results for end-to-end dereverberation network (joint learning) proposed in~\cite{bowu2e2017} is also compared with the proposed work in Table~\ref{table:1}.  

Finally applying the proposed neural model based dereverberation on BF-FDLP baseline (denoted as BF-FDLP + CLSTM derevb. (prop)) yields average relative improvements of $13$\% on the development set and about $4$\% on the evaluation set, compared to the BF-FBANK baseline. After joint training this further improves to $17$\% and $9$\% respectively. The improvement in real condition is much more than that of simulated data. Average relative improvements of $20$\% on the real development set and about $18$\% on the real evaluation set, compared to the BF-FBANK baseline is achieved by the proposed method. This suggests that, even though the jointly learned neural model is trained only with simulated reverberation, it generalizes well on unseen real data. 
\vspace{-0.2cm}
\subsection{CHiME-3 ASR}
The CHiME-3 dataset \cite{chime3} for the ASR has multiple microphone tablet device recording in four different environments, namely, public transport (BUS), cafe (CAF), street junction (STR) and pedestrian area (PED). For each of the above  environments, real and simulated data are present. The real data consists of $6$ channel recordings from WSJ0 corpus sampled at $16$ kHz spoken in the four varied environments. The simulated data was constructed by mixing clean utterances with the environment noise. The training dataset consists of $1600$ (real) noisy recordings and $7138$ (simulated) noisy recordings from $83$ speakers.
\begin{table}[t!]
\caption{Word Error Rate (\%) in CHiME-3 dataset for different features and proposed dereverberation method. Here prop denotes proposed work in this paper.}
\resizebox{\columnwidth}{!}{
\begin{tabular}{@{}l|ccc|ccc@{}}
\toprule
\multicolumn{1}{c|}{\multirow{2}{*}{\textbf{\begin{tabular}[c]{@{}c@{}}Model\\ Features\end{tabular}}}} & \multicolumn{3}{c|}{\textbf{Dev}}                                              & \multicolumn{3}{c}{\textbf{Eval}}                                             \\ \cmidrule(l){2-7} 
\multicolumn{1}{c|}{}                                                                                   & \multicolumn{1}{l}{\textbf{Real}} & \multicolumn{1}{l}{\textbf{Simu}} & \multicolumn{1}{l|}{\textbf{Avg}} & \multicolumn{1}{l}{\textbf{Real}} & \multicolumn{1}{l}{\textbf{Simu}} & \multicolumn{1}{l}{\textbf{Avg}} \\ \midrule
BF-FBANK & 7.8 & 8.0 & 7.9 & 14.0 & 9.7 & 11.8 \\
BF-FDLP  & 7.0 & 8.1 & 7.5 & {12.0} & 10.0 & 11.0 \\
BF-FBANK + CLSTM derevb. (prop)   & 7.2     & 8.3     & 7.7   & 12.9     & 9.8    & 11.4     \\
BF-FBANK + spectral mapping derevb.\cite{spec_map}                                                                                            & 8.0                     & 10.0                      & 9.0                     & 14.3                     & 12.3                      & 13.3                    \\
BF-FBANK + context aware derevb.\cite{santos2018speech}                                                                                            & 7.7                     & 9.9                      & 8.8                     & 13.4                     & 13.3                      & 13.3                    \\
BF-FDLP + CLSTM derevb. (prop)   & 7.2    & 7.9    & 7.5    & 13     & 9.6    & 11.3     \\ 
$~~~~~~~~~~~~~$ + spec. reg. (prop)  & \textbf{6.9 }   & 8.0    & 7.4    & 11.8     & 9.8    & 10.8     \\
$~~~~~~~~~~~~~$ + spec. reg. + joint (prop)  & 7.0   & \textbf{7.7}    & \textbf{7.3}    & \textbf{11.7}     & \textbf{9.3}    & \textbf{10.5}     \\ \bottomrule
\end{tabular}}
\vspace{-0.4cm}
\label{table:2}
\end{table}

\subsubsection{Discussion}
The WER results for experiments on CHiME-3 dataset are shown in Table \ref{table:2}. The FDLP baseline, denoted as BF-FDLP is better than the FBANK baseline (BF-FBANK). We observe  average relative improvements of $8$\% on the development set and about $12$\% on the evaluation set when comparing BF-FDLP and BF-FBANK baseline systems. It can also be seen from Table \ref{table:2} that the proposed dereverberation method improves the FBANK-baseline system. The results based on the implementation of works done by Han et. al.~\cite{spec_map} and Santos et. al.~\cite{santos2018speech} degrade the word error rates further compared to the BF-FBANK baseline. 

In the CHiME-3 dataset, we observed that the significant cause of degradation in the signal quality came from the additive noise sources.  On further investigation, we found that the dereverberation model also resulted in smoothing of the spectral variations in the FDLP spectrogram. In order to circumvent this issue, we regularized the MSE loss with a term that encouraged the spectral channels to be uncorrelated. The regularization parameter was kept at $0.05$. Using this regularized MSE loss, we further improved the BF-FDLP-Dereverberation system results over the dereverberation approach with MSE loss alone. These experiments suggest that even when the audio data does not have significant late reflection components (like CHiME-3 dataset), the proposed approach improves significantly over the baseline method (average relative improvements of $10.3$ \% over the baseline BF-FBANK system in the real dev condition and $23.5$ \% on real eval condtion). 

\subsubsection{VOiCES corpus ASR}

\begin{table}[t!]
\centering
\caption{WER in VOiCES dataset for different features and proposed dereverberation method. Here prop denotes proposed work in this paper.}
\begin{tabular}{l|c|c}
\toprule
\multirow{1}{*}{\textbf{Model Architecture}} & \multicolumn{1}{c|}{\textbf{Dev}}            & \multicolumn{1}{c}{\textbf{Eval}}            
                                      \\ \midrule
BF-FBANK                             & 55.5             & 66.6         \\
BF-FDLP                            & 51.5           & 62.6        \\
BF-FDLP + CLSTM derevb. (prop)                            & 52.8           & 62.4        \\
~~~~~~~~~~~~~ + joint. (prop)                           & \textbf{49.9}           & \textbf{59.8}        \\
\bottomrule
\end{tabular}

\label{table:3}
\end{table}

Since the REVERB challenge dataset  and CHiME-3 dataset are relatively smaller datasets, we wanted to establish the efficacy of the proposed dereverbaration method in a larger dataset. Thus we experimented with VOiCES challenge dataset. VOiCES corpus \cite{voices} is released as part of ``The voices from a distance challenge 2019'' \cite{voices_inter} of Interspeech 2019. For the ASR fixed conditons track, the training set consists of 80-hours subset of LibriSpeech corpus \cite{librispeech}. The training set has close talking microphone recordings from 427 different speakers from quiet environment. The development and evaluation sets consists of 19 hours and 20 hours of distant microphone recordings of varying room, environment and noise conditions.  The significant difference between the training set and development/evaluation set makes the challenge even more difficult. We have used the same acoustic model configurations and hence these results reflect the true acoustic mismatch condition in ASR.

\subsubsection{Discussion}
 The WER results for VOiCES corpus is given in Table. \ref{table:3}. As seen, the baseline FDLP, denoted by BF-FDLP, provides at a better WER compared to the baseline FBANK. denoted as BF-FBANK. This is further improved with joint learning based dereverberation.
 The final WER shows improvement in both development set and evaluation set. A relative WER improvement of $10$\% in both development set and evaluation set over the baseline FBANK system is observed in these experiments.



\section{Analysis}\label{sec:discuss}
In this section the effect of different neural network architectures and various parameters like regularization parameter, $\lambda$, FDLP model order, $p$ on  WER are reported in Tables 4-6 and Figure \ref{fig:bar plot}.
\begin{table}[t!]
\caption{Word Error Rate (\%) in REVERB dataset using different model architectures for dereverberation(without joint training)}\begin{tabular}{@{}l|ccc|ccc@{}}
\toprule
\multirow{2}{*}{\textbf{Architecture}} & \multicolumn{3}{c|}{\textbf{Dev}}              & \multicolumn{3}{c}{\textbf{Eval}}              \\ \cmidrule(l){2-7}
                                     & \textbf{Real} & \textbf{Simu} & \textbf{Avg}   & \textbf{Real} & \textbf{Simu} & \textbf{Avg}   \\ \midrule
BF-FDLP            \cite{anu1}                                                                                & 17.8                     & 6.8                      & 12.3                     & 14.0                     & 7.0                      & 10.5                    \\ \hline 
\multicolumn{7}{c}{Neural Dereverberation} \\  \hline 
3 layer DNN                          & 17.3          & 5.4           & 11.3          & 14.2          & 6.9           & 10.5          \\
5 layer CNN                          & \textbf{16}            & 5.9          & 11         & 13.5          & 7.2           & 10.3          \\
4 layer CNN + 3 layer DNN                        & 17.9          & 5.6           & 11.7          & 14.4          & \textbf{6.7}           & 10.5          \\
7 layer LSTM (1024 units each)             & 17            & \textbf{5.3}           & 11.1          & 14.2          & 7.4           & 10.8           \\
7 layer Resnet                           & 17.4          & 7.9           & 12.6          & 14.8          & 10.3          & 12.5          \\
CNN + DNN + LSTM [2,2,3]                                                                                            & 19.1                     & 6.8                      & 13.0                     & 15.5                     & 7.9                      & 11.7                    \\
CLSTM (4-CNN + 3-LSTM)                                & 16.3 & 5.6  & \textbf{10.9} & \textbf{13.4} & 7.1  & \textbf{10.2}  \\
\bottomrule
\end{tabular}
\label{table:archi}
\end{table}
\subsection{Architecture of Dereverberation Model}
\par Table \ref{table:archi} shows the WER for different neural network architectures. We initially explore a DNN of three feed forward layers. A slight improvement in development set is seen over the FDLP baseline, BF-FDLP. The relative improvement in WER becomes appreciable as we move to 5 layer CNN. The architecture with LSTM alone is promising. We also explore a Resnet \cite{resnet} style architecture which was successful in image recognition. A combination of CNN, DNN and LSTM did not perform well compared to baseline. Finally the CNN + LSTM combination provides the best WER.
\begin{table}[t!]
\caption{WER for various regularization to alleviate spectral smearing CHiME3 dataset. The regularization term is the cross correlation of the spectral bands. In absence of reverberation, alleviating spectral smearing improves the WER.}
\centering
\begin{tabular}{@{}c|ccc|ccc@{}}
\toprule
\multirow{2}{*}{\textbf{Regularizer} \textbf{weight, $\lambda$}} & \multicolumn{3}{c|}{\textbf{Dev}}             & \multicolumn{3}{c}{\textbf{Eval}}             \\ \cmidrule(l){2-7}
                             & \textbf{Real}          & \textbf{Simu} & \textbf{Avg}           & \textbf{Real}          & \textbf{Simu} & \textbf{Avg}           \\ \midrule
$\lambda$ = 0.3                    & 7 & 8.1  & 7.5 & 12.3 & 9.9  & 11.1 \\
$\lambda$ = 0.1                   & 7          & 8.2  & 7.6          & 12.5            & 10  & 11.2          \\
$\lambda$ = 0.05                    & \textbf{6.9}          & \textbf{8}  & \textbf{7.4}         & \textbf{11.8}          & 9.8  & \textbf{10.8}          \\
$\lambda$ = 0.02                    & 7.2          & 8.4  & 7.8         & 12.5          & 10  & 11.2 \\
$\lambda$ = 0                    & 7.2          & 7.9  & 7.5         & 13          & \textbf{9.6}  & 11.3 \\
\cmidrule(l){1-7}
BF-FBANK & 7.8 & 8.0 & 7.9 & 14.0 & 9.7 & 11.8 \\
\bottomrule
\end{tabular}
\label{tab:regu-ch}
\end{table}
\begin{table}[h!]
\caption{WER for various regularization to alleviate spectral smearing REVERB dataset.  The regularization term is the cross correlation of the spectral bands. In the presence of significant reverberation, the extra regularization did not improve the performance.}
\centering
\begin{tabular}{@{}c|ccc|ccc@{}}
\toprule
\multirow{2}{*}{\textbf{Regularizer} \textbf{weight, $\lambda$}} & \multicolumn{3}{c|}{\textbf{Dev}}             & \multicolumn{3}{c}{\textbf{Eval}}             \\ \cmidrule(l){2-7}
                             & \textbf{Real}          & \textbf{Simu} & \textbf{Avg}           & \textbf{Real}          & \textbf{Simu} & \textbf{Avg}           \\ \midrule
$\lambda$ = 0.0                    & \textbf{16.3} & 5.6  & \textbf{10.9} & \textbf{13.4} & 7.1  & \textbf{10.2} \\
$\lambda$ = 0.05                   & 16.6          & \textbf{5.5}  & 11.0          & 14            & 6.9  & 10.4          \\
$\lambda$ = 0.1                    & 16.9          & 5.6  & 11.2         & 13.7          & 6.9  & 10.3          \\
$\lambda$ = 0.2                    & 17.2          & 5.5  & 11.3         & 14.2          & 6.8  & 10.5 \\ 
\cmidrule(l){1-7}
BF-FBANK                                                                                          & 19.1                     & 6.1                      & 12.6                     & 14.7                     & \textbf{6.5}                      & 10.6                    \\
\bottomrule
\end{tabular}
\label{tab:regu-rvb}
\end{table}

\subsection{Spectral Correlation Loss}
As reported in Table~\ref{table:2} on CHiME-3 dataset, an extra loss function which encourages the spectral bands to be uncorrelated improves the ASR performance on noisy data when the data is corrupted by additive noise with minimal reverberation artifacts.  Table \ref{tab:regu-ch}, \ref{tab:regu-rvb} shows the effect of the regularization weight, $\lambda$ on WER in CHiME-3 and REVERB datasets respectively for the spectral correlation loss used in the model learning. The introduction of the spectral correlation loss improves the WER in CHiME-3 dataset. The best results are obtained for a choice of $\lambda = 0.05$.

\par The introduction of spectral correlation loss does not benefit the REVERB challenge dataset.  We hypothesize that this may due to the more dominant effect of temporal smearing seen in the REVERB challenge dataset. For the experiments on the VOiCES corpus, the spectral correlation loss is not used. 

\begin{figure*}[h!]
  \centering
  \includegraphics[width=\textwidth]{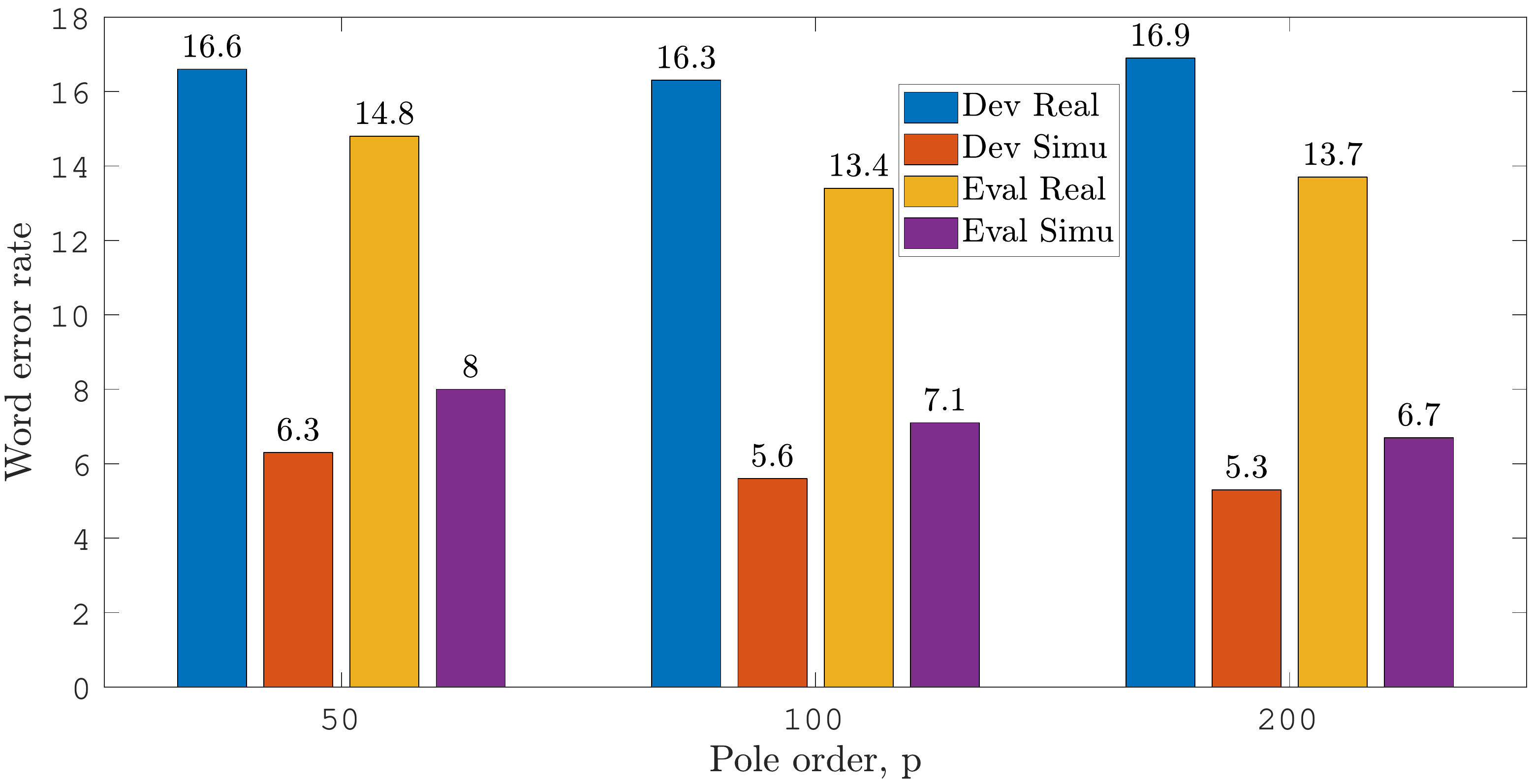}
      \caption{WER (\%) for various model-order, $p$ in FDLP model  for REVERB dataset}
  \label{fig:bar plot}
  \vspace{-0.4cm} 
\end{figure*}
\subsection{Choice of FDLP model order}
\par Figure \ref{fig:bar plot} shows the effect of model order, $p$ used in the FDLP envelope estimation on the WER for the REVERB challenge dataset. The model order $p$ is the number of ``past'' samples used in the auto-regressive modeling of the sub-band DCT signal for a $2$ sec window. While the WER results on the simulated conditions improve with higher model order of the FDLP, the performance on the real conditions is observed to be the best for about $100$ poles per $2$ sec. of audio in each sub-band.
All the other experiments reported in the paper use the $100$ poles per $2$ sec. window of the audio signal. 

\subsection{Discussion on Performance Gains}
All the results reported in Table~\ref{table:1}, Table~\ref{table:2} and Table~\ref{table:3} use a strong baseline system with GEV based beamforming and weighted prediction error (WPE) based enhancement. Hence, we note that all systems use the same pre-processing pipeline and the gains observed over the baseline system are in addition to these enhancement steps. In addition, we also ensure that the baseline FBANK based system, neural enhancement methods explored in the past and the proposed approach have the same sub-band decomposition, feature normalization, acoustic model and language model settings. In this way, the results reported highlight   the effectiveness of the proposed work in suppressing reverberation distortions. 

The methods proposed previously based on neural enhancement and dereverberation improve the performance of the baseline system on the REVERB challenge dataset. However, as seen in Table~\ref{table:2}, in the presence of additive noise conditions on the CHiME-3 dataset, most of these prior works degrade the performance compared to the BF-FBANK baseline system. In this regard, the method proposed in this paper provides significant performance improvements on all three datasets. Further, the results consistently highlight the performance gains of using the joint neural learning framework.  

\section{Summary}\label{sec:summary}
In this paper, we propose a new neural model for dereverberation of temporal envelopes and joint learning of the acoustic model to improve the ASR cost. The joint learning framework combines the envelope dereverberation framework, feature pre-processing and acoustic modeling into a single neural pipeline. This framework is hence elegant and the model can be learned using a joint loss function. Using the proposed neural dereverberation approach and joint learning, we perform speech recognition experiments on the  REVERB challenge dataset as well as on the CHiME-3 dataset. These experiments indicate that the proposed neural dereverberation approach generalizes well on unseen conditions. The analysis of results also highlight the incremental benefits achieved for different choice of hyper-parameters and model architecture settings. The application of the proposed approach for large vocabulary speech recognition experiments on VOiCES dataset further emphasizes the performance benefits.

\section*{Acknowledgment}
This work was funded by grants from Samsung Research India, Bangalore. 

\bibliographystyle{elsarticle-num}
\bibliography{ref}
\end{document}